## Cell-to-cell variability in organelle abundance reveals mechanisms of organelle biogenesis

Sandeep Choubey*[1] Dipjyoti Das[2], Saptarshi Majumdar[1]
[1] Max Planck Institute for the Physics of Complex Systems, Nöthnitzerstraße 38, 01187 Dresden, Germany
[2] Department of Molecular, Cellular and Developmental Biology, Yale University, New Haven, Connecticut, United States of America
* Corresponding Author: sandeep@pks.mpg.de (SC)

### Abstract

How cells regulate the number of organelles is a fundamental question in cell biology. While decades of experimental work have uncovered four fundamental processes that regulate organelle biogenesis, namely, de novo synthesis, fission, fusion and decay, a comprehensive understanding of how these processes together control organelle abundance remains elusive. Recent fluorescence microscopy experiments allow for the counting of organelles at the single-cell level. These measurements provide information about the cell-to-cell variability in organelle abundance in addition to the mean level. Motivated by such measurements, we build upon a recent study and analyze a general stochastic model of organelle biogenesis. We compute the exact analytical expressions for the probability distribution of organelle numbers, their mean, and variance across a population of single cells. It is shown that different mechanisms of organelle biogenesis lead to distinct signatures in the distribution of organelle numbers which allows us to discriminate between these various mechanisms. By comparing our theory against published data for peroxisome abundance measurements in yeast, we show that a widely believed model of peroxisome biogenesis that involves de novo synthesis, fission, and decay is inadequate in explaining the data. Also, our theory predicts bimodality in certain limits of the model. Overall, the framework developed here can be harnessed to gain mechanistic insights into the process of organelle biogenesis.

### Introduction

An organelle is a spatial compartment in eukaryotic cells [1,2] that performs a specialized function. Examples of organelles include vacuoles, Golgi bodies, endoplasmic reticulum etc. Cells tightly regulate organelle number in response to environmental and intra-cellular cues [1–3]. For instance, the number of mitochondria in mammals is tightly regulated in response to their metabolic needs [4]. Yeast cells significantly downregulate vacuole abundance in response to starvation, or upon hypotonic shock [5]. These instances raise the natural question: how is organelle abundance regulated in cells?

Fluorescence microscopy studies of live and fixed cells over the years have led to the discovery of four basic processes that control organelle numbers in cells, namely, de novo synthesis from a pre-existing membrane source [6–8], fission [9–12], fusion [5,13–19], and decay through autophagy or random partitioning during cell division [2,20–22]. For example, mitochondria regulation involves fission and fusion [23]. An important property of organelle biogenesis is that all of the aforementioned processes are inherently stochastic [24,25]. This has led to an alternative approach to unraveling the mechanisms of biogenesis in cells by counting organelle numbers [2,24,26]. Then, the measured steady-state distribution of organelles across a cell population can be used to infer the dynamics of their biogenesis. In a recent study, Mukherjee et al. [24] put forward a general and elegant model of organelle biogenesis consisting of these processes. Using a combination of theory and experiments the authors exploited the cell-to-cell variability in organelle abundance to uncover the kinetic rules of organelle biogenesis. This approach led to newer insights, such as that Golgi body abundance is controlled through the balance of de novo synthesis and decay [24]. While this study along with another follow up work [26] represent the first attempts to uncover the mechanisms of organelle biogenesis using cell-to-cell variability or noise in organelle abundance, these aforementioned studies only looked into the specific limits of the general model. A comprehensive understanding of the impact of different mechanisms underlying organelle biogenesis on the cell-to-cell variability in organelle abundance remains in its infancy.

The goal of this manuscript is to carry out such a systematic exploration. First, we compute the exact analytical expressions for the probability distribution, mean, and variance of the organelle distribution for the general model, and all the different limits of this model. We demonstrate the utility of our theoretical results by applying them to published datasets for peroxisome counts at the single cell level. We show that a proposed mechanism of biogenesis [24,26], where peroxisome number is controlled through de novo synthesis, fission, and decay is inadequate in explaining the data. Moreover, we discover that in a region of parameter space the model predicts bimodality in organelle abundance. Overall, our study provides a general and comprehensive analysis of how different mechanisms of organelle biogenesis control organelle number *in vivo* and illustrates a recipe to extract mechanistic insights from single cell organelle measurements.

**Model**

To shed light on how the mechanisms of organelle biogenesis impact the cell-to-cell variability in organelle abundance across a cell line, we build upon a recent study by Mukherjee et al. [24,26]. In this study, the authors proposed a general model of organelle biogenesis involving four different processes: de novo synthesis, fission, fusion, and decay, as shown in Figure.1A. This phenomenological model essentially combines the hitherto observed mechanisms of biogenesis of various organelles such as mitochondria, vacuole, peroxisome, Golgi body, etc. [6,17]. As shown in Figure.1A, in this model, de novo synthesis of an organelle happens with a zeroth order rate constant $k_d$. Through fission one organelle copy divides and produces two copies at first order rate constant $k_{fis}$ per organelle. Two organelle copies fuse at second order rate constant $k_{fus}$ per organelle squared, and form one organelle copy. Finally, an organelle copy can decay with a first order rate constant $\gamma$ per organelle. It is evident from the model that de novo synthesis and fission increase the number of organelles, while fusion and decay decrease their number (see Figure.1B). Through an interplay between these opposing processes, the organelle abundance reaches a steady-state, whereby the distribution of number of organelles does not change in time. While attaining the steady-state in the different limits of the model is not always guaranteed (for details, see the Materials and methods section), such an assumption has been useful and reasonable in explaining the experimental data [24,26]. In this regime, one can analyze the different limits of the general model and make specific predictions about the steady-state organelle abundance.

For the investigation of the steady-state properties of the model, it is instructive to delineate the above processes [24,26] in the space of organelle number, as shown in Figure.1B. These processes and their corresponding weights allow us to employ a stochastic framework and monitor the time evolution of the organelle number. The probability distribution $P(n, t)$ of having $n$ organelles in a cell at a time $t$ is given by the master equation [24]

$$\begin{aligned}\frac{dP(n,t)}{dt} &= \left[k_d + k_{fis}(n-1)\right]P(n-1,t) + \left[\gamma + k_{fus}n\right](n+1)P(n+1,t) \\ &\quad -\left[k_d + k_{fis}n + \gamma n + k_{fus}n(n-1)\right]P(n,t).\end{aligned} \tag{1}$$

The above equation is an agglomeration of all possible steps that lead to either an increase or decrease in organelle copy number (see the processes in Figure.1B). In principle, the master equation contains all the information about the organelle number distribution and its moments such as the mean and variance etc. However, obtaining exact solutions for the moments and the distribution from this master equation is challenging [24,26]. Alternatively, we make use of the detailed balance condition [26,27] to obtain the steady-state organelle number distribution (see the Materials and methods section). From these steady-state distributions, we compute the mean and variance of organelle numbers using standard functions in *Mathematica.*

**Different mechanisms of organelle biogenesis provide distinct 'fingerprints' in cell-to-cell variability in organelle abundance**

To expound the effect of different mechanisms of organelle biogenesis on the cell-to-cell variability in organelle abundance, we consider different possible limits of the general model. For instance, consider a model of biogenesis that involves non-zero rates of de novo synthesis and decay, while contributions from fission and fusion processes are vanishing; we call this model, de novo synthesis-decay. We identify six such limits of the general model that reach the steady-state: (i) de novo synthesis-decay, (ii) fission-fusion, (iii) de novo synthesis-fusion, (iv) de novo synthesis-fission-decay, (v) de novo synthesis-fusion-decay, and (vi) de novo synthesis-fission-fusion. For a detailed discussion on the conditions of reaching the steady-state for the different possible mechanisms, see the Materials and methods section. Although the (iii) de novo synthesis-fusion and (vi) de novo synthesis-fission-fusion models do have steady-states in terms of organelle number, because of the presence of the de novo synthesis process the organelle size will keep increasing. Hence, these models can be biologically relevant only if there are other cellular mechanisms [22] to maintain mass balance without altering the organelle number (see Materials and methods section for a discussion). It must be noted that we consider only those models which show stationarity in the strict mathematical sense. Clearly, one cannot rule out the possibility of fission-decay or fission-fusion-decay model being biologically relevant, if the fission rate is much greater than the decay rate (for a detailed discussion, see the Materials and methods section). We compute the exact steady-state distribution of organelle numbers, its mean, and variance for each of these six cases as well as the general model (see Table.1). With these results at hand, we perform a comparative analysis of how these different mechanisms affect the cell-to-cell variability in organelle abundance. To carry out such an analysis, throughout the rest of the manuscript we quantify the cell-to-cell variability using a statistical quantity, known as the Fano factor [24,28] which is defined as the ratio of the variance and the mean. We describe the behavior of the Fano factor as a function of the mean as we change the various experimentally tunable rates associated with the four processes.

**De novo synthesis-decay**

The de novo synthesis-decay model is a good starting point owing to its simplicity [24,26]. Abundance of many organelles such as Golgi body [24,29,31–33] is controlled through this mechanism [6,20,29,30]. The number distribution of organelles for this model is characterized by the Poisson distribution (shown in Table.1) [24], which is defined by one effective parameter, given by the ratio of de novo synthesis and decay rate constants. One of the key properties of a Poisson distribution is that its mean and variance are equal i.e., the Fano factor for this model is one, independent of the values of de novo synthesis and decay rate constants, as shown in Fig. 2A. This model thus serves as a good reference point for the comparison of different models, wherein a deviation of the Fano factor from one indicates the presence of other processes. This feature holds the key to our analysis in the rest of the manuscript.

| Mechanism | Steady-state probability distribution | Mean | Fano factor |
|---|---|---|---|
| De novo-decay | $\frac{1}{n!}\lambda^n e^{-\lambda}$, where $\lambda = k_d / \gamma$. | $\lambda$ | 1 |
| Fission-fusion | $\frac{1}{n!}\phi^n\left(\frac{e^{-\phi}}{1-e^{-\phi}}\right)$, where $\phi = k_{fis} / k_{fus}$. $for\ n = 1,2,3....$ $= 0,\ for\ n = 0$ | $\frac{\phi e^{\phi}}{(e^{\phi}-1)}$ | $1-\frac{\phi}{(e^{\phi}-1)}$ |
| De novo-fusion | $\frac{1}{n!(n-1)!}\beta^{n-1/2}\frac{1}{I_1\left(2\sqrt{\beta}\right)}$, $for\ n = 1,2,3....$ $= 0,\ for\ n = 0$ where $\beta = k_d / k_{fus}$. | $\sqrt{\beta}\frac{I_0\left(2\sqrt{\beta}\right)}{I_1\left(2\sqrt{\beta}\right)}$ | $1-\frac{{}_0F_1^R(1,\beta)}{{}_0F_1^R(2,\beta)}+\frac{\beta\ {}_0F_1^R(2,\beta)}{{}_0F_1^R(1,\beta)}$ |
| De novo-fission - fusion | $\frac{1}{\alpha}\frac{\phi^{n-1}}{n!(n-1)!}\frac{\Gamma(\alpha+n)}{\Gamma(\alpha)}\frac{1}{{}_1F_1(1+\alpha,2,\phi)}$, $for\ n = 1,2,3....$ $= 0,\ for\ n = 0$ where $\alpha = k_d / k_{fis}$, $\phi = k_{fis} / k_{fus}$. | $\frac{{}_1F_1(1+\alpha,1,\phi)}{{}_1F_1(1+\alpha,2,\phi)}$ | $\frac{{}_2F_2(\{2,1+\alpha\},\{1,1\},\phi)}{{}_1F_1(1+\alpha,1,\phi)}-\frac{{}_1F_1(1+\alpha,1,\phi)}{{}_1F_1(1+\alpha,2,\phi)}$ |
| De novo-fission - decay | $\frac{q^n(1-q)^{\alpha}}{n!}\frac{\Gamma(\alpha+n)}{\Gamma(\alpha)}$, where $\alpha = k_d / k_{fis}$, $q = k_{fis} / \gamma$. | $\frac{q\alpha}{(1-q)}$ | $\frac{1}{(1-q)}$ |
| De novo-fusion - decay | $\frac{\beta^n}{n!}\frac{\Gamma(\delta)}{\Gamma(\delta+n)}\frac{1}{{}_0F_1(\delta,\beta)}$, where $\beta = k_d / k_{fus}$, $\delta = \gamma / k_{fus}$. | $\frac{\beta\ {}_0F_1^R(1+\delta,\beta)}{{}_0F_1^R(\delta,\beta)}$ | $1-\frac{\beta\ {}_0F_1^R(1+\delta,\beta)}{{}_0F_1^R(\delta,\beta)}+\frac{\beta\ {}_0F_1^R(2+\delta,\beta)}{{}_0F_1^R(1+\delta,\beta)}$ |
| De novo-fission - fusion-decay | $\frac{\phi^n}{n!}\frac{\Gamma(\alpha+n)}{\Gamma(\alpha)}\frac{\Gamma(\delta)}{\Gamma(\delta+n)}\frac{1}{{}_1F_1(\alpha,\delta,\phi)}$ where $\alpha = k_d / k_{fis}, \delta = \gamma / k_{fus}, \phi = k_{fis} / k_{fus}$. | $\frac{\alpha\phi}{\delta}\frac{{}_1F_1(1+\alpha,1+\delta,\phi)}{{}_1F_1(\alpha,\delta,\phi)}$ | $1-\frac{\alpha\phi\ {}_1F_1^R(1+\alpha,1+\delta,\phi)}{{}_1F_1^R(\alpha,\delta,\phi)}+\frac{(1+\alpha)\phi\ {}_1F_1^R(2+\alpha,2+\delta,\phi)}{{}_1F_1^R(1+\alpha,1+\delta,\phi)}$ |

**Table.1.** Formulas for the steady-state probability distributions, as well as for their means and Fano factors for different models of organelle biogenesis are shown. We discuss each of the seven models in the first results section. The different special functions that arise in these different formulas are defined as following: ${}_nF_m(a_1,a_{2,..,}a_n;b_1,b_{2,..,}b_m;z)$ is the generalized hypergeometric function that has *n* parameters of type 1 and *m* parameters of type 2. ${}_nF_m^R(a_1,a_{2,..,}a_n;b_1,b_{2,..,}b_m;z) \equiv \frac{{}_nF_m(a_1,a_{2,..,}a_n;b_1,b_{2,..,}b_m;z)}{\Gamma(b_1)\Gamma(b_2)..\Gamma(b_m)}$ is known as the regularized

hypergeometric function, where $\Gamma(x)$ is the gamma function. Moreover, $I_n(x)$ is the modified Bessel function of the 1st kind of order *n,* where *n* is an integer. For a detailed discussion of special functions please see the references [34,35].

**De novo synthesis-fusion**

The de novo synthesis-fusion model is fully characterized by one effective parameter, defined by the ratio of the rate constants characterizing de novo synthesis and fusion. For this model the steady-state number of organelles never goes to zero. The Fano factor increases as a function of the mean and asymptotically goes to 0.5 when we alter one of the rate constants of this model while keeping the other one fixed, as shown in Figure.2A. While de novo synthesis happens at a constant rate, fusion depends on the square of the number of organelles. Such a reduction in the overall noise level can be comprehended by comparing the de novo synthesis-decay and de novo synthesis-fusion models. These two models differ from each other owing to the manner in which the number of organelles decrease; the reduction in organelle number happens through two different processes, namely decay and fusion. While decay goes linearly with organelle number, fusion depends on the square of the organelle number, as shown in Figure.1B. The relatively rapid change in the "weight" (or, equivalently, in the rate) of fusion as the organelle number fluctuates around its mean value has a strong restoring tendency, thus making the distribution narrower compared to decay. Consequently, the Fano factor becomes less than one. The impact of fusion on noise in organelle numbers has been reported before by Mukherjee et al. [24].

**Fission-fusion**

Next, we consider the Fission-fusion model. In yeast and mammalian cells, the biogenesis of organelles such as vacuole, mitochondria involves both fission and fusion [16,23,36–38]. For the fission-fusion model, we find that the organelle abundance is given by a truncated Poisson distribution, fully characterized by the ratio of the rate constants characterizing fission and fusion. This result is in agreement with a previous theoretical study [26]. It is evident that for this model the number of organelles can never go below one. As shown in Figure.2A, when we tune either one of the two rates of the model, while keeping the other rate fixed, the Fano factor increases as a function of the mean and asymptotically approaches one. In other words, with an increasing mean, the organelle number distribution approaches the Poisson distribution, characteristic of the de novo synthesis-decay model [24,26].

Clearly, the three models corresponding to combinations of two processes lead to distinct predictions for the Fano factor as a function of the mean, as shown in Figure.2A. Since each of these three models depends on one effective parameter, the Fano factor as a function of the mean for each of them is uniquely defined. The results obtained for these models serve as reference points for exploring the models consisting of combinations of three processes.

**De novo synthesis-fission-decay**

De novo synthesis-fission-decay model is defined by two effective parameters (ratio of the de novo synthesis and fission rate constants, and fission and decay rate constants respectively, see Table.1). We seek to explore the behavior of the Fano factor as a function of the mean for this model. This can be achieved in three different ways: i) change the de novo synthesis rate constant keeping the fission and decay rate constants fixed. ii) change the fission rate constant keeping the de novo synthesis and decay rates constants fixed, and iii) alter the decay rate constant keeping the fission and de novo synthesis rate constants fixed. Here we choose the second scenario, motivated by the fact that the fission rate constant of, for instance, peroxisome can be tuned by knocking down genes such as *DNM1*, *VPS1*, etc. The Fano factor versus mean plots for the other two scenarios are shown in the SI (Figure.S1). Using the de novo synthesis-decay model, as our reference, it is evident that the Fano factor increases monotonically as a function of the mean, as shown in Figure.2B. Clearly, the inclusion of the fission process enhances the overall noise, which is in agreement with previous studies [24,26].

**De novo synthesis-fission-fusion**

The de novo synthesis-fission-fusion model is defined fully by two effective parameters (ratio of the de novo synthesis and fission rate constant and fission and fusion rate constants), as shown in Table.1. Like the previous model, we can expound the behavior of the cell-to-cell variability for this model by tuning one of the parameters while keeping the other two parameters constant. As an example, we alter the de novo synthesis rate constant keeping the fission and fusion rate constants fixed, see Figure.2C; rest of the two scenarios are shown in the SI (Figure.S2). Using the fission-fusion model as a reference, we find that as the de novo synthesis rate constant is increased, the de novo synthesis-fission-fusion model predicts an overall lowering of the noise level (see Figure.2B). Moreover, the Fano factor as a function of the mean approaches the de novo synthesis-fusion model. On the other hand, when the fission rate constant is increased keeping the de novo synthesis rate constant fixed, the Fano factor increases as a function of the mean and approaches the curve defining the fission-fusion model. Indeed, the cell-to-cell variability in organelle abundance for this model is bound by the Fano factor-mean curves defining the fission-fusion and de novo synthesis-fusion model.

**De novo synthesis-fusion-decay**

The de novo synthesis-fusion-decay model is characterized by two effective parameters (ratio of the de novo synthesis and fusion rate constants, and fusion and decay rate constants respectively, see Table.1). Fano factor as a function of the mean is plotted by altering the de novo synthesis rate constant while keeping the fusion and decay rate constants fixed, see Figure.2D; For the other possible scenarios, see the SI (Figure.S3). Using the de novo synthesis-fusion model as our reference, we find that the introduction of the decay rate enhances the noise level, as shown in Figure.2D. Moreover, the Fano factor shows a non-monotonic behavior as a function of the mean as we alter the decay rate constant for when the fusion rate constant is much greater than the decay rate constant (see the red curve in Figure.2D). Initially, when the mean is less than one, the Fano factor decreases and shows a minimum around a value of mean equal to one. Subsequently, the Fano factor increases with the mean and asymptotically goes to one. The reason for this non-monotonic behavior is that the fusion process cannot occur when the organelle number is less than two, and hence decay is dominant. On the other hand, at large organelle numbers, the fusion process dominates over the decay process since the fusion rate constant is much greater than the decay rate constant. As a result, the Fano factor of the organelle number switches from a Poissonian behavior (set by the de novo synthesis-decay process) to a behavior set by the de novo synthesis-fusion process (Figure.2D). This cross-over produces the observed non-monotonic behavior in the Fano-factor. For the de novo synthesis-fusion-decay model, the upper-bound of the noise level is set by the noise in a Poisson process, characteristic of the de novo synthesis-decay model, while the lower bound is set by the noise in the de novo synthesis-fusion process. It is evident that the presence of the fusion process decreases noise in organelle abundance which is consistent with previous studies [24,26].

**De novo synthesis-fission-fusion-decay**

Finally, we analyze the general model consisting of all the four processes and explore how these four processes in conjunction impact organelle abundance across a cell population. This model is characterized by three effective parameters (ratio of the de novo synthesis and fission rate constant, fission and fusion rate constant, and fusion and decay rate constant), as shown in Table.1. Using the analytical expressions from Table. 1, we can study the behavior of the Fano factor as a function of the mean by altering one of the parameters of the model while keeping the others fixed. For instance, when we alter the fission rate constant (keeping the other rate constants fixed), the Fano factor shows a non-monotonic behavior as a function of the mean, as shown in Figure.2E. For a small mean level, the Fano factor increases with the mean as we increase the fission rate, which is characteristic of the de novo synthesis-fission-decay model. For higher organelle copy numbers the fission and fusion processes dominate (due to the weights of these processes, see Figure.1B) and the Fano factor asymptotically approaches one, after showing a peak in between. However, as the fusion rate tends to zero, the Fano factor manifests a linear behavior as a function of the mean abundance, characteristic of the de novo synthesis-fission-decay model. For a thorough analysis of the de novo synthesis-fission-fusion-decay model, see the SI (Figure.S4).

Overall, these results imply that we can discern between different mechanisms of organelle biogenesis based on the specific predictions they make for the Fano factor as a function of the mean.

**The de novo synthesis-fission-decay model fails to explain the noise in peroxisome abundance in yeast**

To illustrate the usefulness of our theoretical results, we re-analyze previously published data for single-cell peroxisome counts in *Saccharomyces cerevisiae* [39]. In *Saccharomyces cerevisiae*, peroxisomes are primarily involved in the metabolism of various carbon and nitrogen sources, such as oleic acid, and purines etc. The underlying mechanisms of peroxisome biogenesis remain elusive [40,41]. Various experimental studies have reported the role of de novo synthesis [6,42,43], fission [44,45] and decay [46] in controlling peroxisome number. Most recent studies [24,26] have systematically constructed a model (de novo synthesis-fission-decay) based on these experimental observations and showed that this model captures the essential features of the peroxisome distribution data in glucose and oleic acid-grown yeast cells.

Using our theoretical results, we put this model to test by applying it to single-cell peroxisome abundance measurements [39]. This study by Kuravi et al. [39], seeks to unravel the role of the dynamin-related proteins such as Vps1, Dnm1 etc., in regulating the number of peroxisomes in *Saccharomyces cerevisiae*. The authors deleted various combinations of the genes, *VPS1*, and *DNM1* and noted the change in peroxisome count in glucose and oleic acid media. Interestingly, most of the mutants from this study [39] manifest a Fano factor of less than one (see Figure.3A). As shown before (see Figure.2C, and the SI), if the de novo synthesis-fission-decay model governs the biogenesis of an organelle, its Fano factor always remains equal to or greater than one (see Figure.3A). Hence, the de novo synthesis-fission-decay model of peroxisome biogenesis is inadequate. Our theoretical exploration showed (see Fig.2A) that the Fano factor of organelle number distribution can go below one only if its biogenesis involves fusion. Motivated by this observation, we consider the de novo synthesis-fission-fusion-decay model consisting of all the four processes. To test this model, first, we fit the peroxisome data for the dnm1- vps1 double-deletion yeast strain with the model to find the values of the individual rate constants (see the Materials and methods and SI (Figure.S5)). For simplification, we assume that the fission rate for this strain is vanishing since two of the key fission factors are not present in the cells. Hence, the de novo synthesis-fission-fusion-decay model reduces to the de novo synthesis-fusion-decay model, which is defined by two effective parameters. Having obtained the corresponding parameter values, we are in a position to test the de novo synthesis-fission-fusion-decay model by comparing the predictions this model makes against the data (for details see the Materials and methods). To achieve this goal, we plot the Fano factor as a function of the mean (Figure.3A), as the fission rate is increased; it is assumed here that the fission rate would be higher for the WT strain as well as the single deletion strains that consist of one of the two genes, *DNM1*, *VPS1*, and *FIS1*. Next, we compare these predictions with the peroxisome abundance measurements in yeast cells grown in the glucose and Oleic acid medium. The theory predictions match well with the data for yeast cells grown in the glucose medium. Interestingly, even for cells grown in the oleic acid medium, which is characterized by higher fission rate of peroxisome [47], the data points follow the trend of the theory curve, barring the data point representing the WT yeast (not shown).

In conclusion, the key finding of this section is that the de novo synthesis-fission-decay model, proposed by Mukherjee et al. [47] and later built upon by Craven et al. [26], cannot explain the single-cell peroxisome data, obtained from various different experimental studies.

**Mechanism of organelle biogenesis predicts a bimodal distribution of organelle abundance**

An idea that has gained much traction in recent years is the possibility of achieving phenotypic heterogeneity through non-genetic mechanisms [48]. A large number of studies have focused on how noise and bimodality in gene expression can lead to phenotypic diversification [28,48]. Interestingly, a recent commentary by Chang et al. [25] hints at the possible role of the cell-to-cell variability of organelles in leading to distinct phenotypes. In this light, we theoretically explore the following question: can mechanisms of biogenesis drive bimodality in organelle abundance? In order to answer this query, we look into the full model consisting of all the four processes. We find that when the rate of de novo synthesis is much smaller than the fission rate, the model predicts a bimodal distribution of organelle abundance across a cell population, as shown in Figure.3B. As discussed earlier, cells increase the number of organelles through de novo synthesis and fission. Fission happens when there is at least a single organelle copy in the cell. When the number of an organelle in a cell goes to zero, the cell remains in this state until the organelle number becomes one through a de novo synthesis event. Once the number of the organelle goes to one, the cell can start producing them through fission and the number can quickly increase. Hence, for when the rate of de novo synthesis is much smaller compared to the fission rate, the cells either remain in the zero-organelle state or produces a finite number of organelles, depending on the fusion and decay rate. It must be noted that the fusion and decay rates should be of comparable magnitude for bimodality to arise. While the fission rate should be higher than the rate of fusion and decay for the second mode of the distribution to exist, for when the fission rate is much greater than the fusion and decay rate, the first mode of the distribution corresponding to the zero-organelle state ceases to exist. It must be noted that none of the six limiting models manifest bimodality. It would be interesting to explore the case where abundance of any organelle exhibits bimodality.

## Discussion

Cell-to-cell variability in organelle abundance in a population of cells can be exploited to unravel the governing principles of organelle biogenesis. While the strategy of using cell-to-cell variability to gain mechanistic insights has led to a number of crucial discoveries in different areas of molecular and cellular biology, such as gene expression [49,50] and flagellar rotation [51], it remains less utilized in understanding the regulation of organelle abundance, except for a few studies [2,24,52]. In this manuscript, we explore a general model [24] of organelle biogenesis and explore the relative contributions of different processes associated with this model in controlling the noise in organelle abundance. In order to achieve this goal, we compute the closed-form steady-state organelle number distributions for each of the limiting models of the general model and the corresponding means and variances. Using these analytical results, we show that the change in the Fano factor of an organelle number distribution as a function of the various parameters leads to distinct predictions for the different mechanisms of organelle biogenesis. These specific predictions of the organelle number distribution not only complement the traditional microscopy experiments but also provide a powerful quantitative lens to extract deeper mechanistic insights from them.

We elucidate the utility of our theory by applying it to published data for peroxisome abundance in *Saccharomyces cerevisiae* [39,53]. We show that the de novo synthesis-fission-decay model, proposed in recent studies [24,26], cannot explain single-cell peroxisome counts obtained from other experimental studies [39]. Rather, a general model consisting of all the four processes, namely, de novo synthesis, fission, decay, and fusion can capture the trend of the data. While experimental findings suggest that mature peroxisomes do not fuse in yeast, it remains debated [45,54]. It was suggested [45,54] that the possibility of peroxisome fusion under certain metabolic or environmental conditions cannot be excluded. In spite of the close match between the predictions of the de novo synthesis-fission-fusion-decay model, and peroxisome measurements, we cannot explicitly rule out other possible mechanisms of regulation of peroxisome abundance. For instance, an alternative possibility is that cells control the number of peroxisomes through some feedback mechanism [1]. It was shown in an experimental study [55] that HEX oligomers through a positive feedback mechanism provide a way of controlling composition and abundance of peroxisomes. Also, it has been shown that the rate of peroxisome decay via autophagy depends on the existence of a functional fission pathway [56]. It is possible to extend our model to incorporate some of these findings. For example, in the case of fission-dependent autophagy rates, the decay rate constant can be expressed as a function of the fission rate constants. More importantly, deviations from this simple model could actually pave the way for discovering the abovementioned mechanisms of organelle abundance control, such as the coupling between the different processes affecting organelle copy number, organelle-size dependent rates or feedback [2] etc. Nonetheless, we make use of the de novo synthesis-fission-fusion-decay model as a simple scenario that explains the data and provides mechanistic insights into the process of peroxisome biogenesis that leads to experimentally testable predictions.

We apply our theory to peroxisome data to demonstrate the utility of our theory; nevertheless, our modeling framework is rather general in spite of its simplicity and is not limited to peroxisome. The processes this model incorporates have been experimentally observed in the context of biogenesis of various organelles such as vacuole, Golgi body, mitochondria, etc. [24,26].

One key question our manuscript deals with is whether cell-to-cell variability in organelle abundance plays any functional role [25]. While many studies to date have identified molecular-level variability such as gene expression noise as an important source of phenotypic heterogeneity [28,48], it's not clear if organelle-level heterogeneity can play a role, if any, in creating phenotypic variability [25]. In a recent commentary, Chang et al. [25] hypothesized a possible role of cell-to-cell organelle variability in the context of disease, in particular, if some diseases exhibit more or less variability compared to non-diseased states [57]. Our analytical results concretize this hypothesis by making specific and experimentally testable predictions for generating bimodality in organelle abundance, where the two modes of the distribution can signify two different phenotypes. It must be noted that other scenarios where fission/fusion/decay depend on organelle composition and size can also potentially lead to bimodality.

In conclusion, we have provided a theoretical framework and related analytical tools to analyze single-cell experiments that produce organelle number distribution, to extract information about the dynamics of organelle biogenesis in cells. The combination of single-cell experiments and theory holds the promise of uncovering comprehensive kinetic information about the process of organelle biogenesis.

## Materials and methods

### On the conditions of reaching the stationary-state

The general model, as shown in Figure.1A consists of four processes, namely, de novo synthesis, fission, fusion, and decay. While de novo synthesis and fission increase the number of organelles, fusion, and decay decrease their number, as shown in Figure.1B. Through an interplay between these opposing processes, the distribution of organelle abundance becomes stationary. Evidently, any possible combination of these processes should at the least consist of one process that increases organelle number and one process that decreases organelle number. We can in principle construct four limiting models that are combinations of two processes: de novo synthesis-decay, fission-fusion, de novo synthesis-fusion, fission-decay. However, out of these four combinations, the fission-decay model does not have any steady-state. When the fission rate constant is much greater than the decay rate constant, the number of organelles tends to keep growing. On the other hand, when the rate constant of decay is greater than the fission rate constant, the organelle number eventually goes to zero. The cell remains in this state since for fission to increase the number of organelles, the cell needs to have at least one organelle copy. Hence the fission-decay model does not lead to a steady-state where the organelle number is finite.

For combinations of three processes, there are four possible limiting models: de novo synthesis-fission-fusion, de novo synthesis-fusion-decay, de novo synthesis-fission-decay, fission-fusion-decay. Amongst these limiting models, we note that the de novo synthesis-fission-decay process reaches a steady-state only when the fission rate constant is lesser than the decay rate constant. Otherwise, the organelle number keeps growing. When the decay rate constant is greater than the fission rate constant, the organelle number does not get frozen at the zero-organelle state due to the presence of de novo synthesis. Thus, the inclusion of de novo synthesis to the fission-decay model leads to a non-trivial steady-state. This same line of argument shows that a combination of fission, fusion, and decay cannot also reach a steady-state in absence of de novo synthesis. All other combinations of the processes, as mentioned above, naturally lead to steady-state conditions for any choice of the rates. Along the same lines, it can be argued that the fission-fusion-decay model doesn't have a non-trivial steady-state.

In this paper, we consider those models which show stationarity in the strict mathematical sense. Let us consider the fission-decay model. For this model, once the system reaches the zero organelle-state it remains there forever. Hence in the limit of the decay rate constant being higher than the fission rate constant, there exists a trivial steady-state. However, if the fission rate constant is much higher than the decay rate constant, the system may never go to the zero-organelle state. The same goes for the fission-fusion-decay model. Hence one cannot rule out the possibility of fission-decay or fission-fusion-decay model being biologically relevant if the rate constant of the fission process is much greater than the decay rate constant.

**Detailed balance**

Here we employ the detailed balance condition following ref. [26,27] to obtain the steady-state organelle number distribution. To justify the applicability of the detailed balance condition, we follow the same line of argument as ref. [58]. Let us consider two organelle number-states, *i* and *j*. Moreover, let $J_{ij}$ denote the steady-state probability current between two states *i* and *j*, given by $J_{i->j} = P(i)W_{i->j} - P(j)W_{j->i}$, where *P*(*i*) is the probability of having *i* number of organelles. Here $W_{i->j}$ is the transition rate from the *i*-th to the *j*-th state. The state space for the organelle numbers does not have any loops because the numbers go linearly from 0, to 1, from 1 to 2, and so on. Correspondingly, the steady-state is characterized by a single constant probability current *J*. Furthermore, because *P*(*N*) tends to zero for large *N*, we must have *J* = 0. Thus, all probability currents vanish in the steady state. Hence, at the steady-state, the detailed balance condition would imply that the frequency of transition from a state of *n* organelle copies to the state of *n*-1 organelle copies must equal the frequency of transition from *n*-1 organelle copy state to *n* organelle copy state, where n= {0, 1, 2,…}. For instance, if we consider the de novo synthesis-decay model, the detailed balance condition would imply the following mathematical condition,

$$k_d P(n-1) = \gamma n P(n).$$

Here the probability of having *n*-1 and *n* organelles in the cell is given by *P*(*n*-1) and *P*(*n*) respectively. The rate of de novo synthesis is given by $k_d$, and $\gamma$ is rate of decay. This recursion relation allows us to find a relationship between the probability of having *n* organelles *P*(*n*), and zero organelles *P*(0) in the cell respectively, which is given by

$$P(n) = \frac{1}{n!}\left(\frac{k_d}{\gamma}\right) P(0).$$

It is evident that *P*(0) has to be evaluated for obtaining the exact expression for *P*(*n*). To achieve this goal, we use the normalization condition that

$$\sum_{n=0}^{\infty} P(n) = 1,$$
$$\sum_{n=0}^{\infty} \frac{1}{n!}\left(\frac{k_d}{\gamma}\right)^n P(0) = 1.$$

We can evaluate this sum using *Mathematica* and get a final expression for *P*(*n*), which is given by

$$P(n) = \frac{1}{n!}\left(\frac{k_d}{\gamma}\right)^n \exp\left(-\frac{k_d}{\gamma}\right) .$$

Similarly, the distribution of organelle abundance for all the different limits of the general model as well the full model can be obtained. Using standard functions in *Mathematica*, we can also obtain the mean and variance of the organelle distribution for all the models. We have also attached a Mathematica file as a supplementary material. The *Mathematica* file can also be found on Github: https://github.com/schoubey123/Analytical_Calculations.

**Limitations of the model**

The model of organelle biogenesis we explore here is an 'effective' model, where we assume the different rates to be constant. It is potentially a simplistic assumption as some of these rates can depend on the size of the organelles i as the fusion rate of peroxisomes [59]. While it is possible to incorporate such size dependence, we believe that the model considered here provides the simplest scenario and hence provide the null-predictions. Any deviation from these models would hint at the presence of other more complicated mechanisms such as organelle-size dependent rates or feedback [2]. Two of the limiting models we consider in our analysis, (iii) de novo synthesis-fusion and (vi) de novo synthesis-fission-fusion models do have steady-states in terms of organelle number, but the mass of organelles would keep increasing indefinitely. Clearly, these models can be biologically relevant only if there are other cellular mechanisms such as possible membrane removal, etc. [22] to maintain mass balance without altering the organelle number. While we analyze these models for the sake of completeness, the obvious next step would be to consider size dependent rates to explore how the model predictions change. This would allow us to also garner a clear understanding of how cells regulate the number and composition of organelles [1].

Some of the models we consider do not include decay, and the reduction in number of organelles happens through fusion. One would imagine that in cells, organelles would encounter some form of decay on account of the various cellular processes. However, if the decay rate constant is much smaller than the rate constant of fusion, then organelle abundance will still be primarily dictated by fusion.

**Parameter extraction**

We evaluate the utility of our analytical results by applying them to published data to gain mechanistic insights into the biogenesis of peroxisome. To this end, we have reanalyzed peroxisome single-cell count data. We extract the peroxisome data from the published peroxisome number distribution plots in Fig. 1(A-J) of ref. [39] by using DigitizeIt, a free online tool for digitizing data plots. From this distribution we compute the mean and Fano factor for the various mutant strains, as shown in Figure.3A.

Next, in order to make the Fano factor-mean prediction plot for de novo synthesis-fission-fusion-decay model (see blue curve in Figure.3A), we fit the data for the dnm1- vps1 double-deletion yeast strain with the de novo synthesis-fusion-decay model to find the values of the individual rates. Here we assume that the fission rate for this strain is vanishing since two of the known fission factors are not present in the cells. It must be noted that we cannot uniquely determine the kinetic rate constants associated with the different processes defining the de novo synthesis-fusion-decay model since the distribution of organelle abundance corresponding to different mechanisms depends on the ratios of the parameters. Hence, we need to set the value of one of the rate constants to one and measure the other rate constants with respect to that parameter. The value of the decay rate constant is set to one. We extract the following parameters: $k_{fus}$=36.39 $t^{-1}$, $k_d$=3.46 $t^{-1}$, and $\gamma$=1 $t^{-1}$, $k_{fis}$=0 $t^{-1}$.

Error bars in Fano factor (Figure.3A) represent the standard deviation of 1000 independent, resampled data sets obtained using the method of Bootstrapping in Matlab.

**Acknowledgement**

We wish to thank Tylor Herman and Sakunatala Chatterjee for detailed comments on the manuscript. SC would like to thank Shankar Mukherjee for his initial inputs.

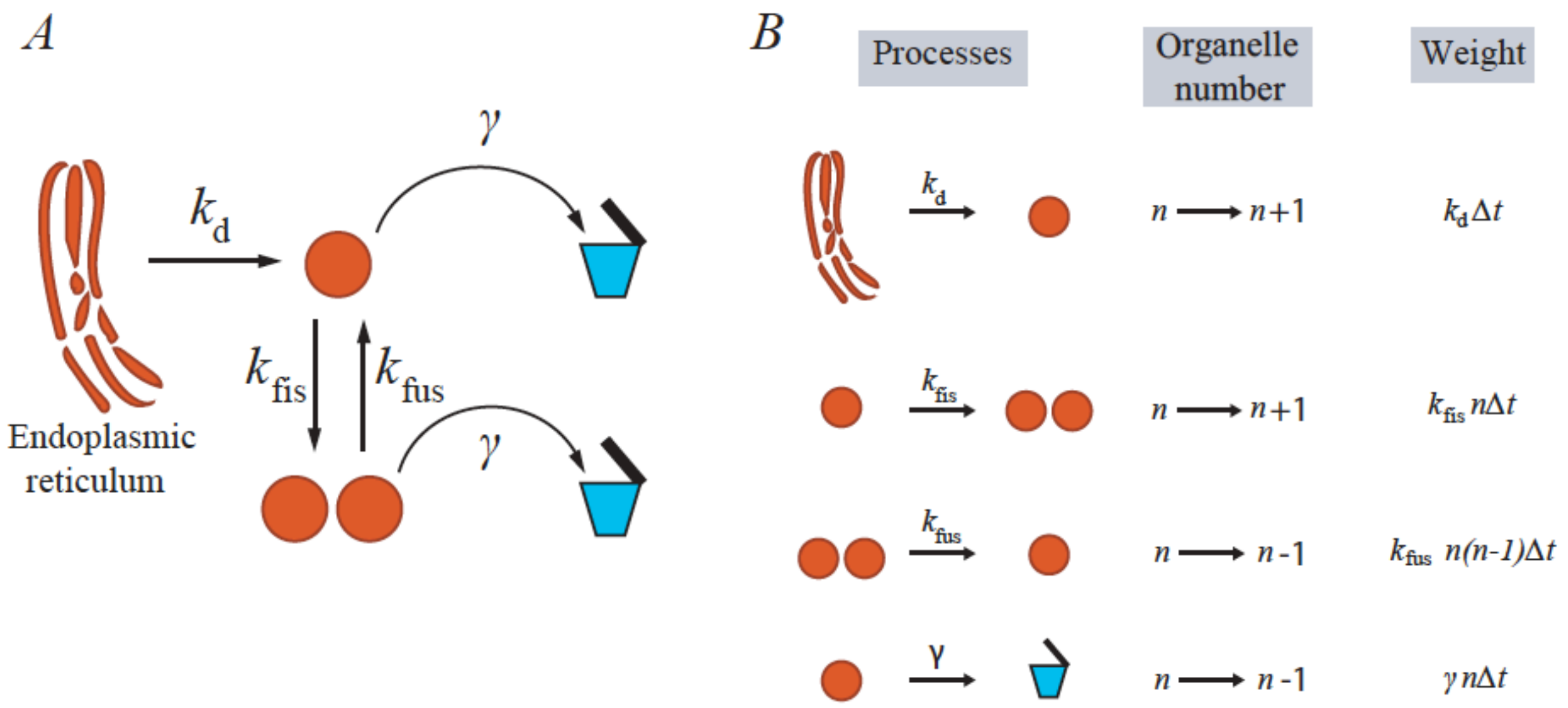

**Figure.1. Model of organelle biogenesis.** A) Organelle biogenesis involves four mechanisms, namely, de novo synthesis, fission, fusion, and decay. The probability per unit time of de novo synthesis is given by zeroth order rate constant $k_d$. $k_{fis}$ is the first order rate constant characterizing fission, $k_{fus}$ is the second order rate constant defining fusion, and $\gamma$ is the first order rate constant defining decay, as proposed by Mukherjee et al. [24]. From this model, we compute the probability distribution of organelle abundance as well as its mean and variance. B) List of possible reactions leading to either an increase or decrease in the organelle number and their respective weights. The weights constitute the probability that each reaction will occur during a time interval, $\Delta t$.

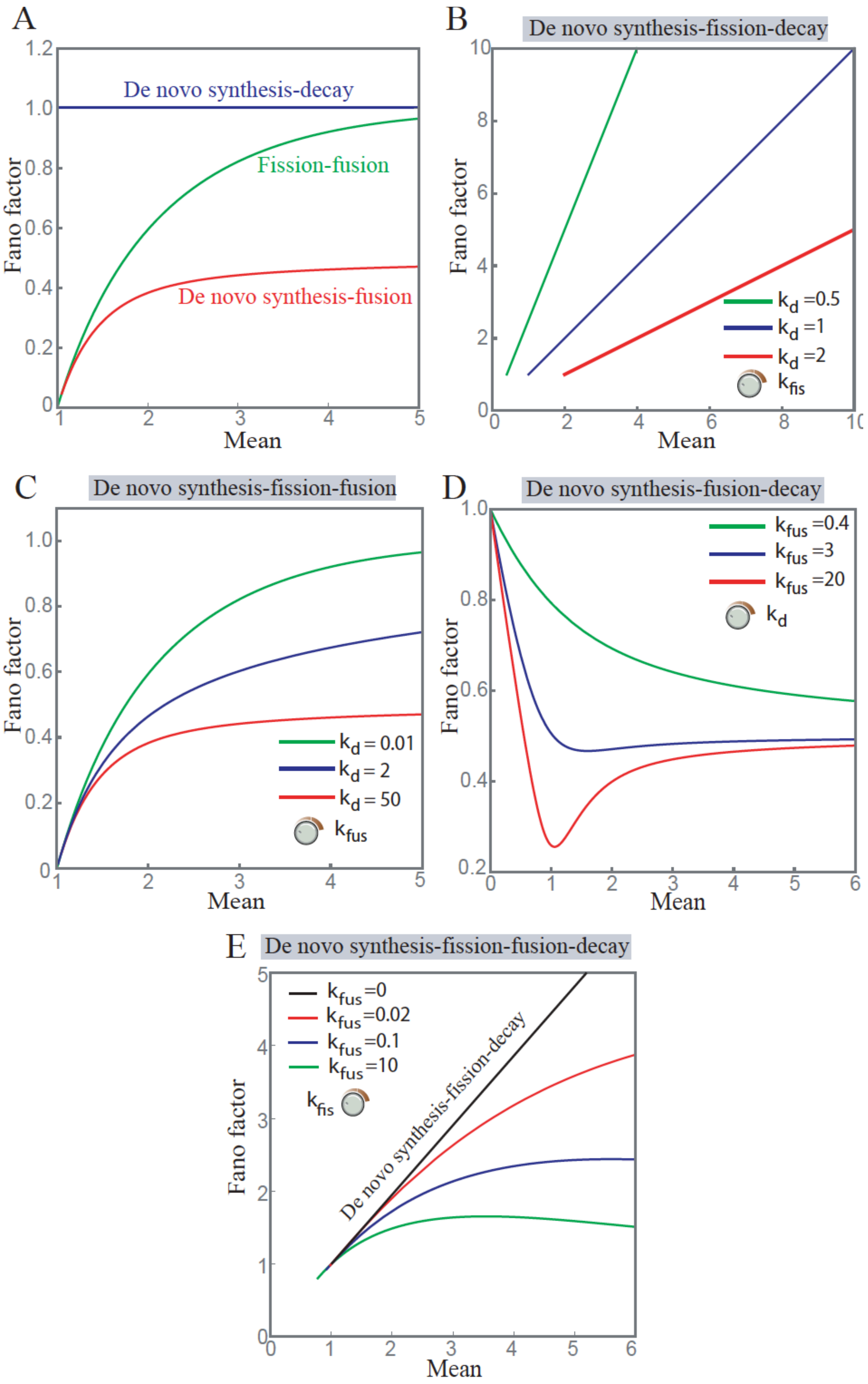

**Figure.2. Noise profiles for different models of organelle biogenesis**. Using the detailed balance condition [26,27], we computed the Fano factor of the organelle number distribution for the different models of organelle biogenesis. (A) Three different models of biogenesis involving two processes: de novo synthesis-fusion (red), fission-fusion(green), and de novo synthesis-decay (blue). These models lead to qualitatively distinct predictions for the Fano factor as a function of the mean. (B) De novo synthesis-fission-decay: We tune the fission rate constant while keeping the other rates fixed to generate the plots. The family of curves corresponds to the different values of $k_d$, where we keep the other rates fixed, $\gamma$=1 $t^{-1}$. (C) De novo synthesis-fission-fusion: Rate of fission is tuned to generate the plots, while we keep the other rate constants fixed. The family of curves correspond to the different values of $k_d$; fusion rate is given by $k_{fus}$ = 1 $t^{-1}$. (D) De novo synthesis-fusion-decay: The de novo synthesis rate is tuned to make the Fano factor versus mean plots. The family of curves corresponds to the different values of $k_{fus}$. The decay rate is given by $\gamma$=1 $t^{-1}$. E) In this plot, we vary $k_{fis}$, while keeping $k_d$ and γ constant (Value of both the rate constants have been kept as one) and explore how the noise profile evolves over different values of $k_{fus}$. When $k_{fus}$=0, the curve becomes a straight line which is characteristic for the de novo synthesis-fission-decay.

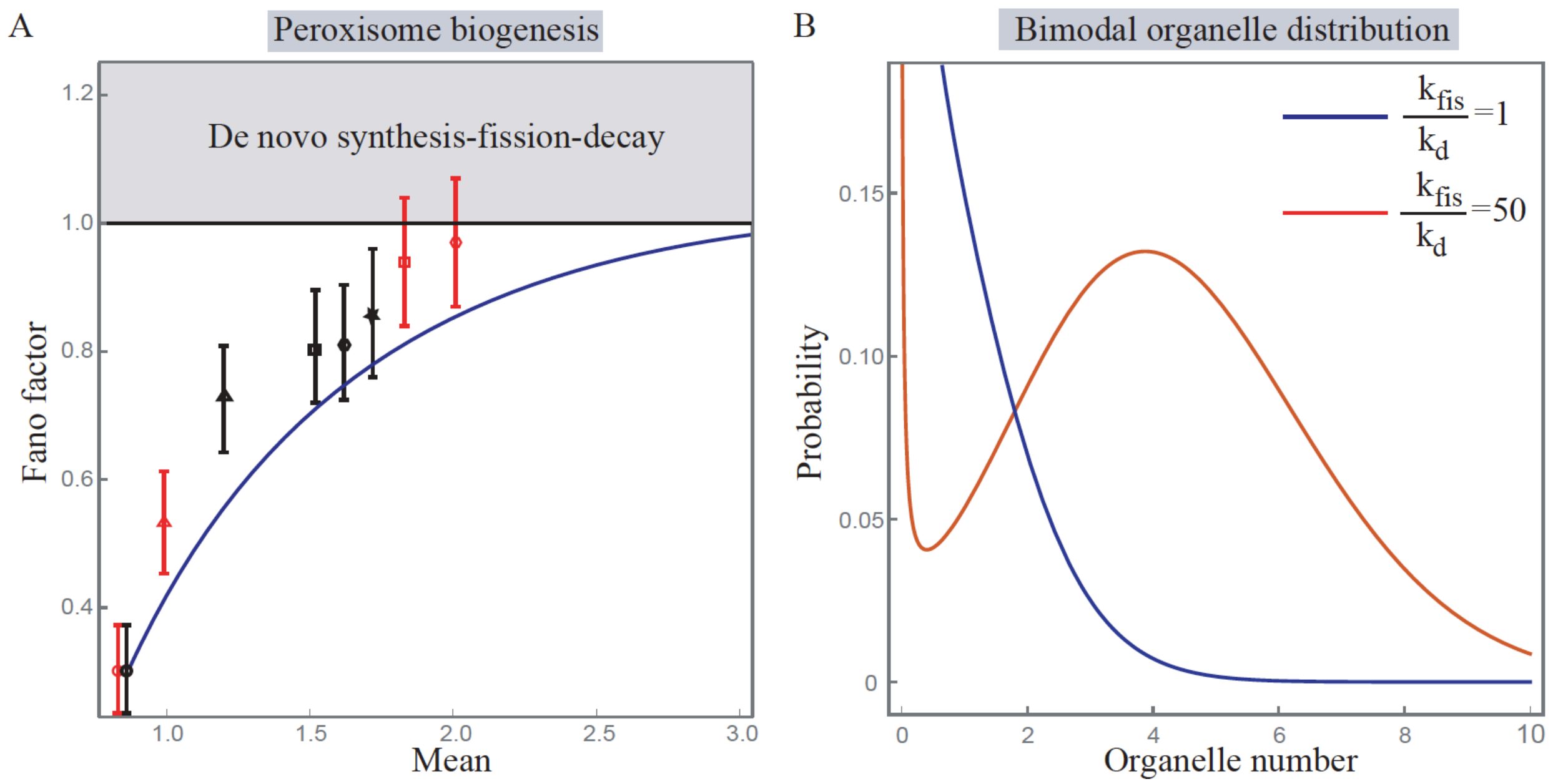

**Figure.3.** (A) **Peroxisome biogenesis:** The black data points represent the experimentally measured Fano factor of peroxisome number distribution as a function of the mean for various *Saccharomyces cerevisiae* mutant strains, grown in glucose medium, with combinations of deleted *VPS1* and *DNM1*, taken from the ref [39]. The data points represent the following mutants: i) o *vps1Δdnm1Δ*, ii) Δ *vps1Δ*, iii) □ *fisΔ* iv) ◊ WT, v) ∗ *dnm1Δ*. The red data points represent the same yeast strains, grown in the oleic acid medium. Corresponding data points are given by i) o *vps1Δdnm1Δ*, ii) Δ *vps1Δ*, iii) □ *fisΔ*, iv) ◊ *dnm1Δ*. The blue curve shows the de novo synthesis-fission -fusion-decay model prediction for how the Fano factor changes as a function of the mean when we alter the fission rate constant while keeping the other rate constants fixed. The other rate constants are given by $k_{fus}$=34.3 $t^{-1}$, $k_d$=3.45 $t^{-1}$, and $\gamma$=1 $t^{-1}$. (B) **Bimodality in organelle abundance:** The steady-state number distributions of organelles are shown for different values of the de novo synthesis rate constant for the de novo synthesis-fission-fusion-decay model. When the rate of de novo synthesis is much greater than the fission rate, the organelle distribution manifests bimodality. The rate constants corresponding to the blue curve are $k_d$=0.1 $t^{-1}$, $k_{fis}$=0.1 $t^{-1}$, $k_{fus}$=1 $t^{-1}$, and $\gamma$=1 $t^{-1}$. The rate constants corresponding to the red curve are $k_d$=0.1 $t^{-1}$, $k_{fis}$=5 $t^{-1}$, $k_{fus}$=1 $t^{-1}$, and $\gamma$=1 $t^{-1}$.

## Supplementary Information

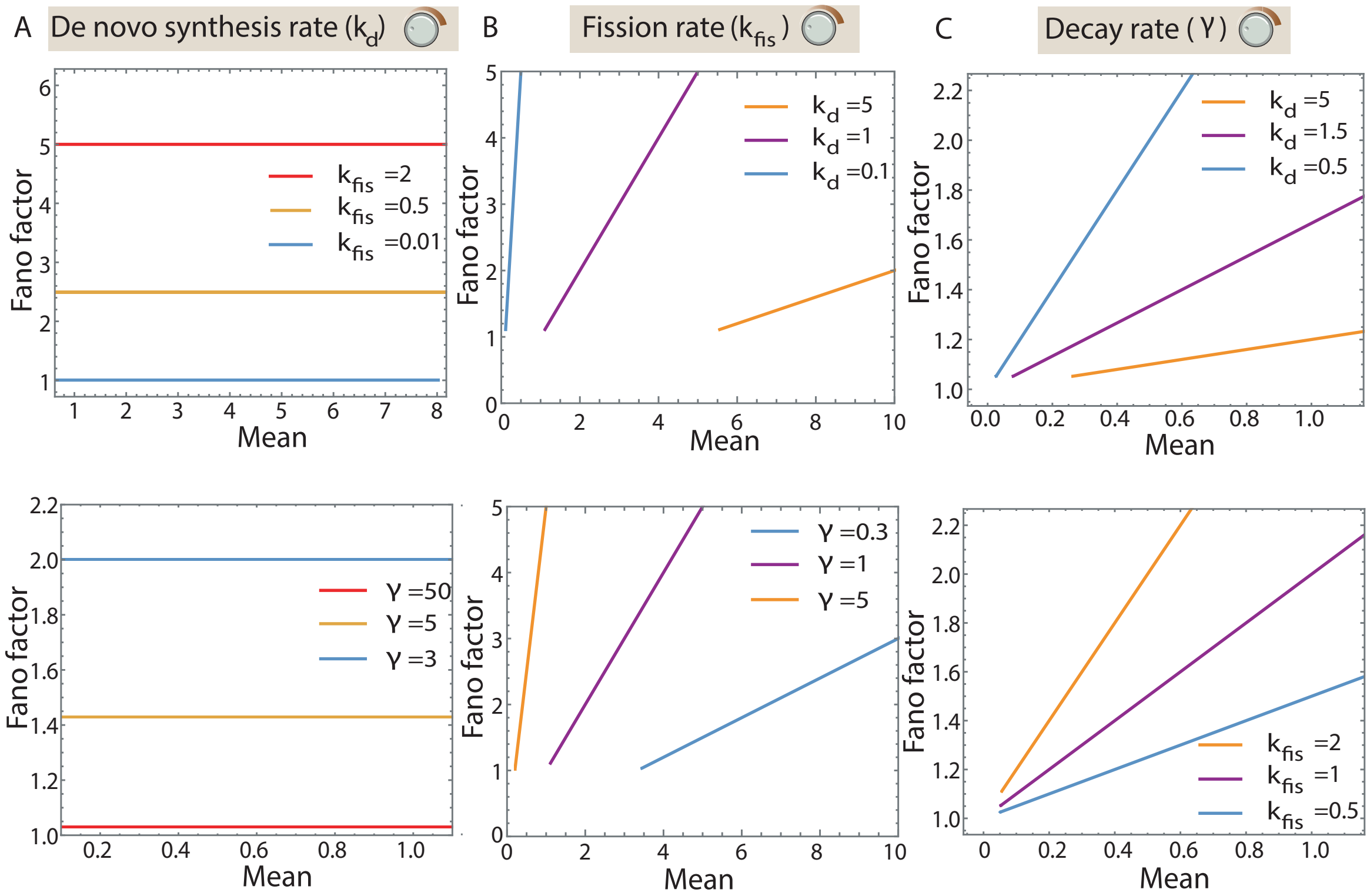

**Figure S1. Noise profile for the de novo synthesis-fission-decay model.** We plot Fano factor vs mean curves by tuning de novo synthesis rate($k_d$)in (A), fission rate constant ($k_{fis}$) in (B) and decay rate constant($\gamma$) in (C). For the rest of the two parameters, we keep one of them constant and take three different values of the other parameter. The values of the constant are – A.1) $\gamma$ = 2.5, A.2) $k_{fis}$= 1.5 ; B.1) $\gamma$ = 1 ,B.2)$k_d$ = 1 ; C.1) $k_{fis}$= 1 ,C.2)$k_d$ = 1. Thus, we obtain a family of curves for each plot which helps us to understand how the noise profile evolves with the various rate constants. The different parameters and values are mentioned in the plots themselves. As it can be observed from the plots, the Fano factor never goes below one. When we vary $k_d$, the Fano factor remains constant with variation of mean, and when we tune the other two parameters, it linearly increases.

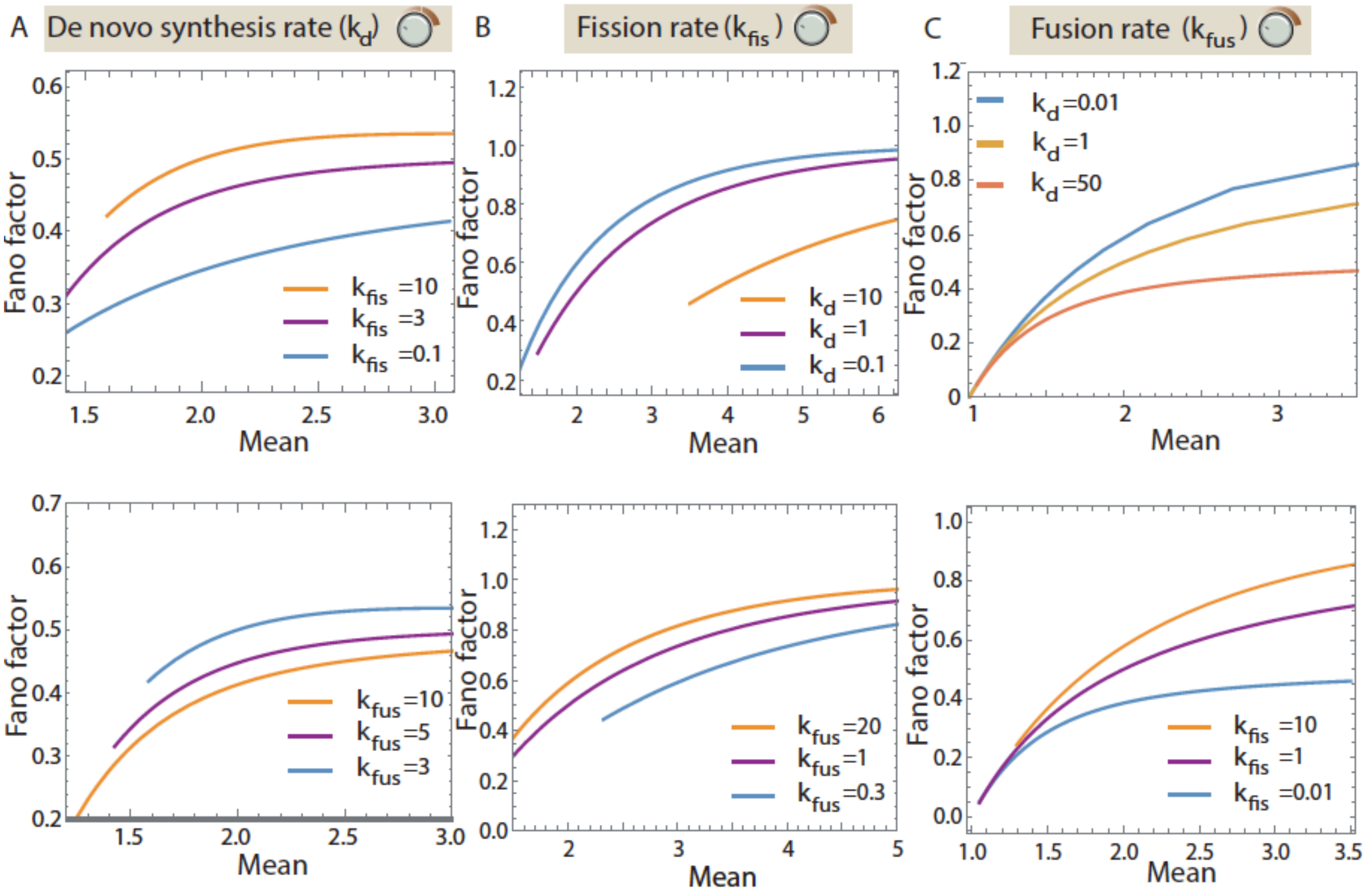

**Figure S2. Noise profile for the de novo synthesis-fission-fusion model.** We plot Fano factor vs mean curves by tuning de novo synthesis($k_d$) in (A), fission rate constant ($k_{fis}$) in (B) and fusion rate constant ($k_{fus}$) in (C). For the rest of the two parameters, we keep one of them constant and take three different values of the other parameter. The values of the rate constants are – A.1) $k_{fus}$ = 5 , A.2)$k_{fis}$ = 3; B.1) $k_{fus}$ = 1, B.2) $k_d$ = 1 ; C.1) $k_{fis}$ = 1, C.2) $k_d$ = 1. We obtain a family of curves for each plot. The different parameters and values are shown. The Fano factor always lies below one.

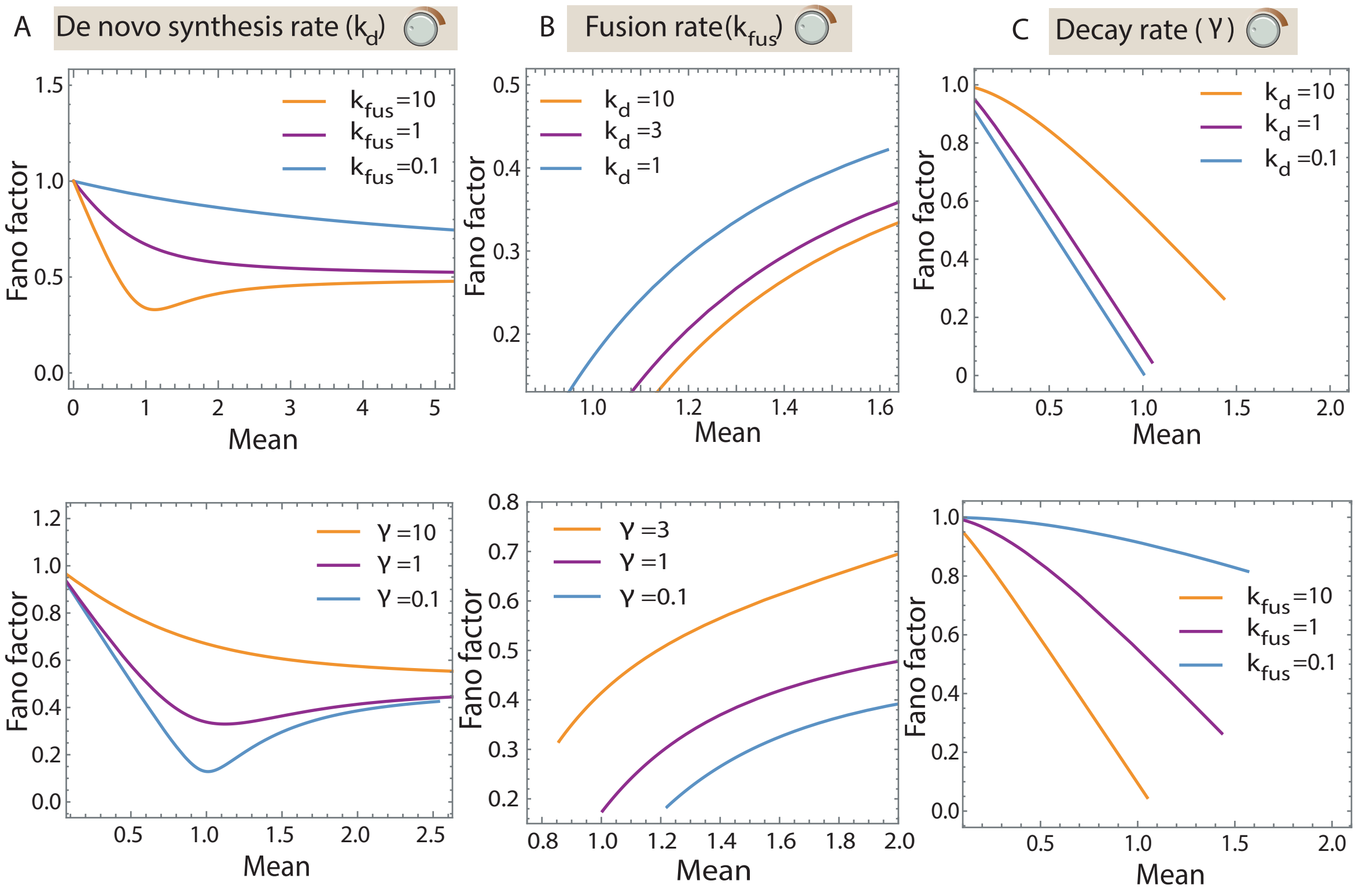

**Figure S3. Noise profile for de novo synthesis-fusion-decay model.** We plot the Fano factor vs mean curves by tuning de novo synthesis($k_d$) in (A), fusion rate constant ($k_{fus}$) in (B) and decay rate constant($\gamma$) in (C). For the two of the other parameters of the model, we keep one of them constant and take three different values of the other parameter. The values of the rate constants are – A.1) $\gamma = 1$, A.2) $k_{fus} = 10$; B.1) $\gamma = 0.1$, B.2) $k_d = 10$ ; C.1) $k_{fus} = 10$, C.2) $k_d = 1$. Different parameter values are shown in the plots. As it can be observed from the plots, fano factor A) decreases, then increases when $k_d$ is varied, B) non-linearly increases when $k_{fus}$ is varied and C) non-linearly decreases when $\gamma$ is varied.

373

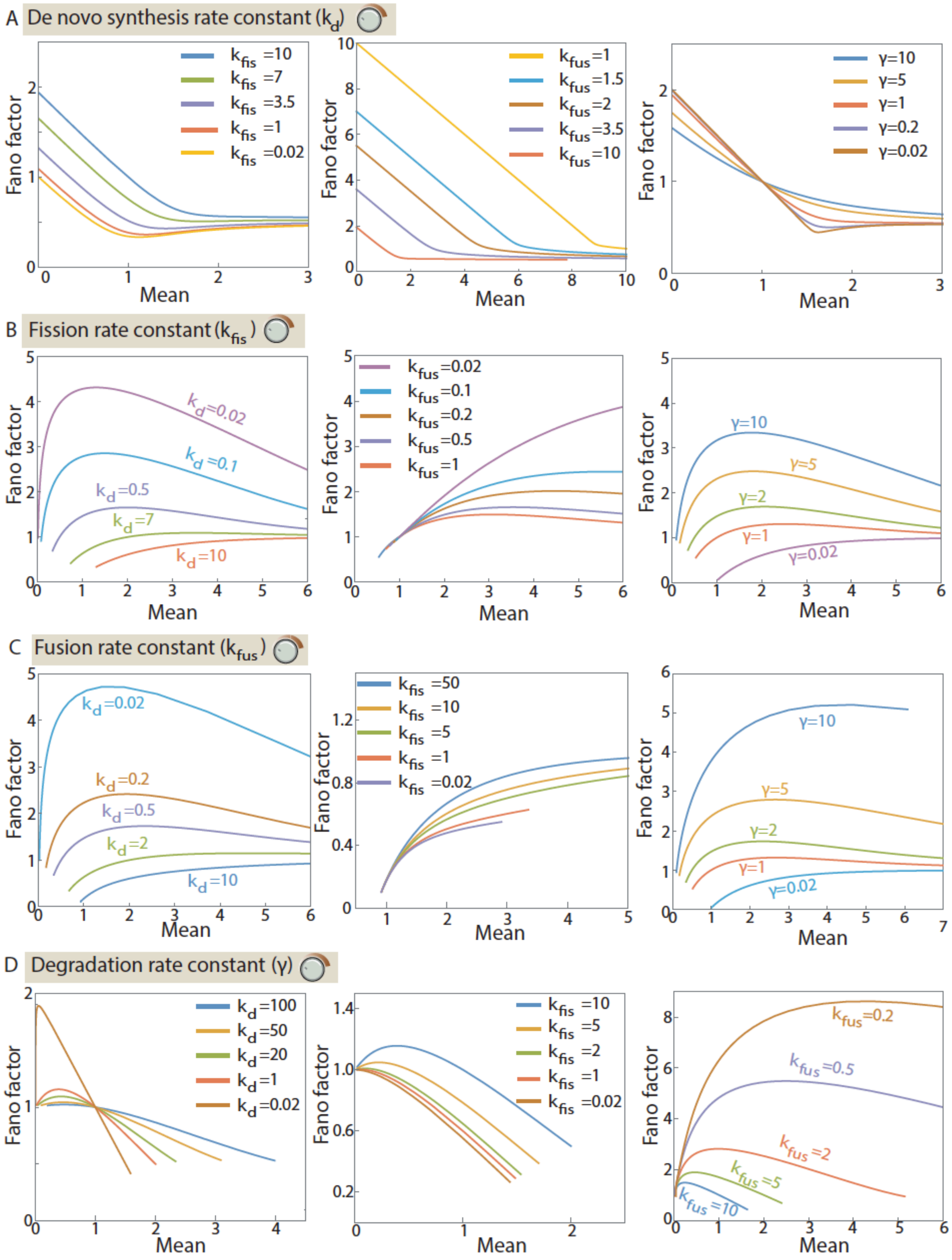

374
375
376 **Figure S4. Noise profiles for organelle biogenesis when all the four mechanisms are present:** We plot
377 the fano factor vs mean curve when all the four processes (de novo synthesis, fission, fusion and decay)

are involved and inspect all the cases. For the plots of one particular row, we tune one of the rate constants – de novo synthesis rate constant ($k_d$) in the 1st row, fission rate constant ($k_{fis}$) in 2nd row, fusion rate constant ($k_{fus}$) in 3rd row and decay rate constant (γ) in the last row. we plot for different values of the other rate constants to check how they behave in different regimes, which gives us sets of family of curves. The values of the various rate constants are – A.1) $\gamma = 1$, $k_{fus} = 10$ , A.2) $\gamma = 1$ , $k_{fis} = 10$ , A.3) $k_{fis} = 10$, $k_{fus} = 10$. B.1) $\gamma = 1$ , $k_{fus} = 10$ ,B.2) $\gamma = 1$ , $k_d = 1$ , B.3) $k_d = 1$ , $k_{fus} = 10$. C.1) $\gamma = 1$, $k_{fis} = 10$ , C.2) $\gamma = 1$, $k_d = 10$ , C.3) $k_d = 1$, $k_{fis} = 20$. D.1) $k_{fis} = 10$, $k_{fus} = 10$ , D.2) $k_d = 10$ , $k_{fus} = 10$ , D.3) $k_d = 1$ , $k_{fus} = 10$. The name and values of the rate constant that is varied to generate a particular family of curves is given in each plot. From this diagram, we can – a) observe how does the concavity/shape/slope of a curve evolves under variation of a parameter, b) differentiate and categorize the regimes where noise is greater than/less than/ equal to one.

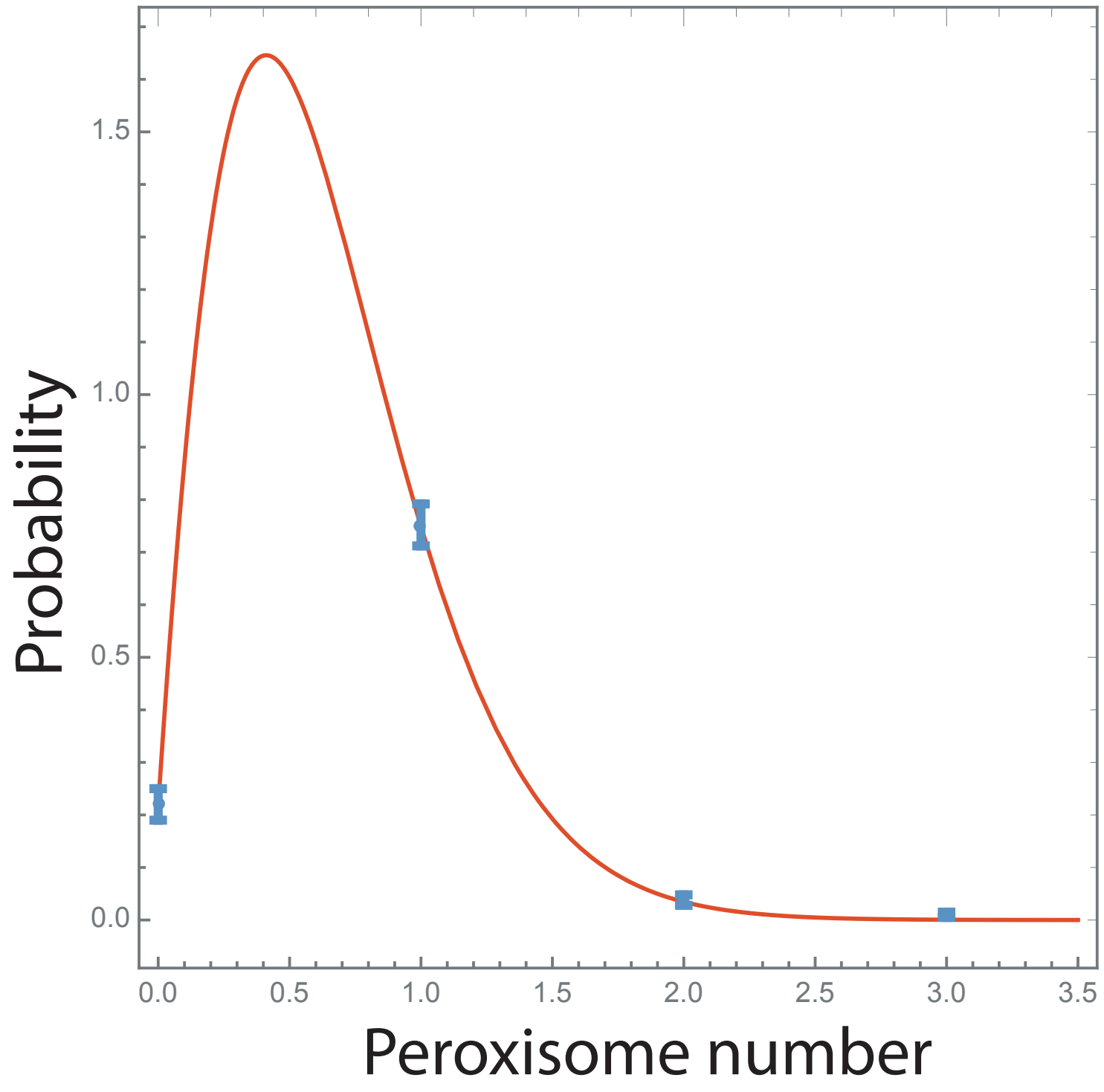

**Figure S5.** We fit the data for the dnm1- vps1 double-deletion yeast strain with the de novo synthesis-fusion-decay model to find the values of the individual rates. The value of the decay rate is set to one. We extract the following parameters: $k_{fus}$=36.4 $t^{-1}$, $k_d$=3.45 $t^{-1}$and $\gamma$=1 $t^{-1}$, $k_{fis}$=0 $t^{-1}$. To demonstrate the goodness of the fit, we have done the Kolmogorov-Smirnov test[1], a well-known tool in statistics. Kolmogorov-Smirnov test is a goodness-of-fit test with the null hypothesis $H_0$ that the sample points have been taken from a certain probability distribution and an alternative hypothesis $H_\alpha$ which counters the former one. The test returns a probability value p, which, if small, denotes that the sample points have a lesser probability of belonging to that particular distribution. Here, we use the test on the peroxisome data points with our predicted distribution, which returns a p-value of 1, suggesting that the model fits the data well. The test has been performed in Mathematica with the help of the KolmogorovSmirnovTest function.